\documentclass[sigconf]{acmart}

\pdfoutput=1

\usepackage{glossaries}
\usepackage{makecell}
\usepackage{multirow}
\usepackage{pifont}
\usepackage{siunitx}
\usepackage[caption=false]{subfig}

\copyrightyear{2025}
\acmYear{2025}
\setcopyright{cc}
\setcctype{by}
\acmConference[CF '25]{22nd ACM International Conference on Computing Frontiers}{May 28--30, 2025}{Cagliari, Italy}
\acmBooktitle{22nd ACM International Conference on Computing Frontiers (CF '25), May 28--30, 2025, Cagliari, Italy}\acmDOI{10.1145/3719276.3725172}
\acmISBN{979-8-4007-1528-0/2025/05}

\newacronym{sota}{SOTA}{state-of-the-art}
\newacronym{fpga}{FPGA}{field programmable gate array}
\newacronym{asic}{ASIC}{application-specific integrated circuit}
\newacronym{fub}{FUB}{functional unit block}
\newacronym{vv}{V\&V}{validation and verification}
\newacronym{gpp}{GPP}{general purpose processor}

\newacronym{hpc}{HPC}{high performance computing}
\newacronym{ml}{ML}{machine learning}
\newacronym{isa}{ISA}{instruction set architecture}
\newacronym{fp}{FP}{floating-point}
\newacronym{dl}{DL}{deep learning}
\newacronym{la}{LA}{linear algebra}
\newacronym{ip}{IP}{intellectual property}
\newacronym[firstplural=systems-on-chip (SoCs)]{soc}{SoC}{system-on-chip}
\newacronym{mpsoc}{MPSoC}{multi-processor system-on-chip}
\newacronym[firstplural=networks-on-chip (NoCs)]{noc}{NoC}{network-on-chip}
\newacronym{hw}{HW}{hardware}
\newacronym{sw}{SW}{software}
\newacronym{swapc}{SWaP-C}{space, weight, power, and cost}
\newacronym{mcp}{MCP}{multi-core processor}
\newacronym{rr}{RR}{round-robin}
\newacronym{vm}{VM}{virtual machine}
\newacronym{vpn}{VPN}{virtual page number}
\newacronym{ppn}{PPN}{physical page number}
\newacronym{lsb}{LSB}{least significant bit}

\newacronym{mac}{MAC}{multiply-accumulate}
\newacronym{fem}{FEM}{finite element analysis}
\newacronym{simd}{SIMD}{single-instruction, multiple-data}
\newacronym{rtl}{RTL}{register transfer level}
\newacronym{dlt}{DLT}{data layout transform}

\newacronym{fifo}{FIFO}{first in, first out}
\newacronym{fu}{FU}{functional unit}
\newacronym{alu}{ALU}{arithmetic logic unit}
\newacronym{fpu}{FPU}{floating-point unit}
\newacronym{ssr}{SSR}{stream semantic register}
\newacronym{issr}{ISSR}{indirection stream semantic register}
\newacronym{tcdm}{TCDM}{tightly-coupled data memory}
\newacronym{dma}{DMA}{direct memory access}
\newacronym{sm}{SM}{streaming multiprocessor}
\newacronym{vlsu}{VLSU}{vector load-store unit}
\newacronym{dsa}{DSA}{domain-specific accelerator}
\newacronym{ha}{HA}{hardware accelerator}
\newacronym{fsm}{FSM}{finite state machine}
\newacronym{llc}{LLC}{last-level cache}
\newacronym{d2d}{D2D}{die-to-die}
\newacronym{dram}{DRAM}{dynamic random access memory}
\newacronym{tid}{TID}{transaction ID}
\newacronym{spm}{SPM}{scratchpad memory}
\newacronym{dspm}{DSPM}{data scratchpad memory}
\newacronym{ispm}{ISPM}{instruction scratchpad memory}
\newacronym{tlb}{TLB}{translation lookaside buffer}
\newacronym{plru}{PLRU}{pseudo least-recently-used}
\newacronym{csr}{CSR}{control and status register}
\newacronym{cam}{CAM}{content-addressable memory}
\newacronym{mmu}{MMU}{memory management unit}

\newacronym{os}{OS}{operating system}
\newacronym{vmm}{VMM}{virtual machine monitor}
\newacronym[longplural={page table entries}]{pte}{PTE}{page table entry}
\newacronym{isr}{ISR}{interrupt service routine}

\newacronym{spvv}{SpVV}{sparse vector-vector multiplication}
\newacronym{spmv}{SpMV}{sparse vector-matrix multiplication}
\newacronym{spmm}{SpMM}{sparse matrix-matrix multiplication}
\newacronym{csrmv}{CsrMV}{CSR matrix-vector multiplication}
\newacronym{csrmm}{CsrMM}{CSR matrix-matrix multiplication}

\newacronym{csf}{CSF}{compressed sparse fiber}
\newacronym{csc}{CSC}{compressed sparse columns}
\newacronym{bcsr}{BCSR}{blocked compressed sparse rows}

\newacronym{axi4}{AXI4}{Advanced eXtensible Interface 4}
\newacronym{amba}{AMBA}{Advanced Microcontroller Bus Architecture}
\newacronym{sram}{SRAM}{static random-access memory}
\newacronym{vmsa}{VMSA}{virtual memory system architecture}
\newacronym{pmsa}{PMSA}{protected memory system architecture}

\newacronym{wcet}{WCET}{worst-case execution time}
\newacronym{rtunit}{REALM unit}{real-time regulation and traffic monitoring unit}
\newacronym{mtunit}{M\&R unit}{monitoring and regulation unit}
\newacronym{cps}{CPS}{cyber-physical system}
\newacronym{crtes}{CRTES}{critical real-time embedded system}
\newacronym{heicps}{He-iCPS}{heterogeneous integrated cyber-physical system}
\newacronym{ecu}{ECU}{electronic control unit}
\newacronym{mcs}{MCS}{mixed criticality system}
\newacronym{ima}{IMA}{integrated modular avionics}
\newacronym{adas}{ADAS}{advanced driver assistance system}
\newacronym{axirealm}{AXI-REALM}{AXI real-time regulation and traffic monitoring}
\newacronym{mpam}{MPAM}{memory system resource partitioning and monitoring}
\newacronym{dos}{DoS}{denial of service}
\newacronym{hwrot}{HWRoT}{hardware root of trust}

\begin{document}


\title{CVA6-VMRT: A Modular Approach Towards Time-Predictable Virtual Memory in a 64-bit Application Class RISC-V Processor}

\author{Christopher Reinwardt}
\email{creinwar@iis.ee.ethz.ch}
\orcid{0009-0004-3184-0763}
\affiliation{%
    \institution{Integrated Systems Laboratory}
    \city{ETH Zurich}
    \country{Switzerland}
}

\author{Robert Balas}
\authornote{Both authors contributed equally to this research.}
\email{balasr@iis.ee.ethz.ch}
\orcid{0000-0002-7231-9315}
\affiliation{%
    \institution{Integrated Systems Laboratory}
    \city{ETH Zurich}
    \country{Switzerland}
}

\author{Alessandro Ottaviano}
\authornotemark[1]
\email{aottaviano@iis.ee.ethz.ch}
\orcid{0009-0000-9924-3536}
\affiliation{%
    \institution{Integrated Systems Laboratory}
    \city{ETH Zurich}
    \country{Switzerland}
}

\author{Angelo Garofalo}
\email{angelo.garofalo@unibo.it}
\orcid{0000-0002-7495-6895}
\affiliation{%
    \institution{Department of Electrical, Electronic, and Information Engineering}
    \city{University of Bologna}
    \country{Italy}
}

\author{Luca Benini}
\email{lbenini@iis.ee.ethz.ch}
\orcid{0000-0001-8068-3806}
\affiliation{%
    \institution{Integrated Systems Laboratory}
    \city{ETH Zurich}
    \country{Switzerland}
}
\affiliation{%
    \institution{Department of Electrical, Electronic, and Information Engineering}
    \city{University of Bologna}
    \country{Italy}
}

\renewcommand{\shortauthors}{Reinwardt et al.}
\begin{abstract}

The increasing complexity of autonomous systems has driven a shift to integrated heterogeneous SoCs with real-time and safety demands. 
Ensuring deterministic WCETs and low-latency for critical tasks requires minimizing interference on shared resources like virtual memory. 
Existing techniques, such as software coloring and memory replication, introduce significant area and performance overhead, especially with virtualized memory where address translation adds latency uncertainty.
%
%
To address these limitations, we propose \emph{CVA6-VMRT}, an extension of the open-source RISC-V CVA6 core, adding hardware support for predictability in virtual memory access with minimal area overhead. \emph{CVA6-VMRT} features dynamically partitioned Translation Look-aside Buffers (TLBs) and hybrid L1 cache/scratchpad memory (SPM) functionality. It allows fine-grained per-thread control of resources, enabling the operating system to manage TLB replacements, including static overwrites, to ensure single-cycle address translation for critical memory regions.
Additionally, \emph{CVA6-VMRT} enables runtime partitioning of data and instruction caches into cache and SPM sections, providing low and predictable access times for critical data without impacting other accesses.
In a virtualized setting, CVA6-VMRT enhances execution time determinism for critical guests by 94\% during interference from non-critical guests, with minimal impact on their average absolute execution time compared to isolated execution of the critical guests only.
%
%
This interference-aware behaviour is achieved with just a 4\% area overhead and no timing penalty compared to the baseline CVA6 core.

\end{abstract}



\maketitle

\section{Introduction}
\label{sec:intro}


With the growing adoption of \glspl{adas} and autonomous driving, automotive platforms face increased computational demand on \glspl{ecu}. 
Addressing this demand through a federated system by adding separate \glspl{ecu} for each task is impractical, as it significantly impacts the size, weight, power, and cost (SWaP-C) metrics. 
Therefore, there has been a transition from a federated, physically split approach to heterogeneous \glspl{mcs}, allowing multiple tasks to be executed concurrently on a single platform~\cite{MCKINSEY_AUTOMOTIVE_SURVEY,AUTOMOTIVE_PREDICTABLE}.

When multiple timing-critical tasks are consolidated onto an integrated \gls{soc}, maintaining execution time predictability becomes challenging due to the interference in shared resources, such as the processor, interconnects, and main memory. 
In this work, we consider inter-task interference within a single processor core.
In particular, we focus on interference in the virtual-to-physical memory address translation process, which introduces uncertainty in memory access times and the variability caused by state-dependent cache latency.

Consider a scenario where a real-time task and a general-purpose task run in separate \glspl{vm} sharing a single processor core through time-division multiplexing. If the general-purpose task, such as rendering passenger information, is memory-intensive, the \glspl{tlb} and caches will likely store translations and data related to this \gls{vm}. When a timing-critical interrupt for the real-time task occurs, it is probable that neither the required \gls{isr} nor the corresponding \glspl{pte} are cached, necessitating page table miss handling and potentially main memory accesses. This process is particularly costly in a virtualized environment, as each intermediate guest-level page table access requires a complete address translation at the hypervisor level. Consequently, the response time of the \gls{isr} is influenced by the processor’s non-deterministic execution history, leading to wide \gls{wcet} estimates and inefficient hardware utilization.

%
Approaches to reduce this overhead include physical isolation on a processor core level by assigning tasks to separate processor cores depending on their criticality~\cite{cortex_r82_blog}. As the critical tasks are run in isolation from the other tasks, they do not suffer from interference in core-local resources. Other work reduces \gls{tlb} interference by coloring the virtual addresses used in the task to limit the available \gls{tlb} resources~\cite{7108391}.
The approach in~\cite{tricore_mmu} uses a custom virtual memory architecture, employing fully software-managed \glspl{tlb}. The decision of which virtual-to-physical translations need to be cached and which can be replaced is therefore taken by supervisor software, enabling arbitrarily complex replacement schemes.
To reduce the non-determinism of memory access latency due to caches, the addition of separate software-managed core-local \glspl{spm} provides the possibility to place critical code and data close to the processor core~\cite{tricore_mmu,cortex_r82_ref_manual,microchip_polarfire_soc_ref_manual, gaisler_leon5_ref_manual}. 

However, these measures come with drawbacks. Using a custom virtual-memory architecture with software-managed \glspl{tlb} or virtual address coloring adds complexity to the \gls{os} and, in the former case, increases the number of page-fault interrupts and context switches. Adding additional processor cores or \gls{spm} memory macros causes significant area overhead for hardware resources that might be underused.

%
To address the interference challenges, in this work, we propose minimal hardware extensions and demonstrate them on an open-source application class RISC-V processor design.
The extensions expose control over the conventionally hidden shared resources of \glspl{tlb} and cache memory. By controlling the replaceable \gls{tlb} entries, supervisor software can create isolated \gls{tlb} states for different tasks. Furthermore, by dynamically changing the ratio of cache and \gls{spm} memory, supervisor software can choose the appropriate amount of \gls{spm} space without wasting area.

\subsubsection{Contribution}
%
We present the following contributions: 
\begin{enumerate}


    \item We design a software-configurable hardware mechanism to dynamically partition (section~\ref{sec:tlb_part}) and lock (section~\ref{sec:tlb_lock}) \gls{tlb} entries to prevent interference in the \glspl{tlb}, thus ensuring constant and predictable memory address translation times in the presence of multiple competing virtual machines.

    \item We design a runtime-configurable hardware unit that partitions the instruction and data caches into cache and \gls{spm} regions (section~\ref{sec:spm}), ensuring fast and predictable memory access for critical \glspl{vm}. This approach allows \gls{spm} resources to be tailored to application requirements with minimal area overhead.

    \item We integrate the extended processor core into a mixed-criticality platform and evaluate it in a virtualized environment hosting guests of different criticality, focusing on interference mitigation for a critical guest. Our experiments (section~\ref{sec:eval}) demonstrate a \emph{94\%} reduction in execution time variability of the critical guest under interference from non-critical \glspl{vm}, with minimal impact on the mean execution time, reclaiming most of the predictability of execution of the critical \gls{vm} in isolation. To assess the cost of our changes, we synthesized the design in a \SI{16}{\nano\meter} technology node, demonstrating no timing impact and an area overhead of only \emph{4\%} with respect to the baseline core.
    
\end{enumerate}

\section{Architecture}
\label{sec:arch}

\begin{figure}[t]
    \centering
    \includegraphics[width=\columnwidth]{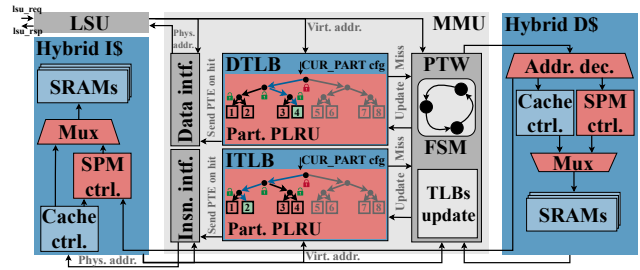}
    \caption{%
        CVA6-VMRT MMU and hybrid cache/SPM subsystem. Modified architectural blocks are highlighted in red.
    }
    \Description{CVA6-VMRT MMU and hybrid cache/SPM subsystem. Modified architectural blocks are highlighted in red.}
    \label{fig:archi:cva6vmrt}
\end{figure}

Our approach to increasing the predictability of critical tasks focuses on the memory system of the processor core, namely the virtual-to-physical memory translation and the core-local cache architecture of the open-source processor core CVA6 (formerly Ariane)~\cite{ariane}. Figure~\ref{fig:archi:cva6vmrt} provides an overview of the \gls{mmu} and cache subsystem in \emph{CVA6-VMRT}, where we highlight in red modified or new architectural blocks.

By dynamically controlling the available \gls{tlb} resources (section~\ref{sec:tlb_part}) through the \gls{plru} tree, we eliminate inter-task interference. Furthermore, by providing protected, software-defined overwrites of \gls{tlb} entries (section~\ref{sec:tlb_lock}), we enable software to specify cached \glspl{pte} directly.
Additionally, by adding the option to trade cache for \gls{spm} space at runtime (section~\ref{sec:spm}), \emph{CVA6-VMRT} allows the \gls{os} to trade-off between average-case performance and predictability of the access time to critical code and data. The following sections detail the implementation of the proposed extensions.

\subsection{PLRU TLB replacement}
\label{sec:tlb_part}

\subsubsection{Vanilla CVA6 PLRU policy}
A \gls{tlb} caches virtual-to-physical translations to avoid the cost of a page table walk for every virtual memory access. CVA6 has two dedicated level-1 \glspl{tlb} for instructions and data.
The \glspl{tlb} in CVA6 are implemented as fully associative register-based caches, storing the recently used \glspl{pte}.
As the number of available \gls{tlb} entries is limited, some entries are evicted from a fully occupied \gls{tlb} by hardware once a new entry needs to be cached.
CVA6 uses a \gls{plru} approach for this decision, aiming to evict no longer needed entries while preserving recently used \glspl{pte}.
To efficiently track the usage history of \gls{tlb} entries, CVA6 uses a binary-tree-based \gls{plru} implementation. When a new \gls{tlb} entry is required in a full \gls{tlb},
the eviction logic traverses the \gls{plru} tree from the root to a leaf node, following the currently \emph{chosen} branch at each internal node. The reached leaf node is then replaced
by the new entry, and all chosen edges on the used path are flipped away from the newly added node.
Similarly, on every \gls{tlb} hit, the edges on the path towards the entry containing the hitting \gls{pte} are pointed away from this entry to maximize its lifetime.

\subsubsection{CVA6-VMRT PLRU partitioning mechanism}
To gain control over the available \gls{tlb} resources, we modify the \gls{plru} replacement scheme by adding constraints on each node in the \gls{plru} tree, dictating which edges (if any)
are available for traversal. These constraints are specified in a custom \texttt{CUR\_PART} \gls{csr} as a bitmap, where each bit corresponds to one partition. A partition refers to a power of two of \gls{tlb} entries, given by the total number of \gls{tlb} entries divided by the number of partitions, which can be parameterized. If a bit in the \texttt{CUR\_PART} \gls{csr} is set, the corresponding \gls{tlb}
resources are modifiable and protected if their associated bit is clear.
By modifying this bitmap, the \gls{os} or hypervisor can dynamically select the amount of \gls{tlb} entries accessible to running tasks.
If different tasks get assigned non-overlapping partitions, \emph{CVA6-VMRT} ensures complete separation of \gls{tlb} replacements and isolates the \gls{tlb} states
against interference from other tasks sharing the same processor core.

In addition to the \texttt{CUR\_PART} \gls{csr} tracking the currently enabled partitions, we add two auxiliary control registers to aid in switching between different configurations quickly. The additional registers are \texttt{LAST\_PART} and \texttt{RESTORE\_LAST\_PART}.
Whenever the \texttt{CUR\_PART} \gls{csr} is written, the \texttt{LAST\_PART} register is updated with the value in \texttt{CUR\_PART} before the write. To restore the \texttt{LAST\_PART} state again into the \texttt{CUR\_PART} register, a write to the \texttt{RESTORE\_LAST\_PART} register with the \gls{lsb} set suffices.
The \texttt{LAST\_PART} and \texttt{RESTORE\_LAST\_PART} registers are beneficial during context switching as they help to save and restore the \gls{tlb} partition bitmap used by the interrupted task.
This allows \texttt{CUR\_PART} to be overwritten with the first instruction in an interrupt handler, minimizing the handler's impact on the previous task's \gls{tlb} state.
If the supervisor handler determines to return to a different task with a differing set of active partitions, overwriting the \texttt{LAST\_PART} \gls{csr} with the new value as part of the task switch lets the trap handler restore the new state into the \texttt{CUR\_PART} register.

\subsection{CSR-based TLB locking}
\label{sec:tlb_lock}
In conjunction with control over the replacement scheme, we also provide direct control over a parameterizable amount of \gls{tlb} entries, using a combination of three \glspl{csr} per locked entry.
This allows the \gls{os} or hypervisor to statically define \gls{tlb} entries, preventing their replacement. The content of these software-provided \gls{tlb} entries does not necessarily need to exist in the system's page tables, giving the management software full flexibility. The \gls{os} has to provide (i) the \gls{vpn} and flags of the mapping, (ii) the leaf \gls{pte} for the translation, and (iii) the ASID or VMID value used for this mapping using three \glspl{csr}.

Once all three registers have their valid bit set, the \gls{tlb} entry associated with this locking slot is marked as unreachable in the \gls{plru} tree, and its contents are provided by the \glspl{csr}.
This way, we can ensure a constant latency for virtual-to-physical memory translations covered by the locked entry, which is helpful for hard real-time tasks with strict response time requirements, as the memory translation latency is made small and predictable.

\begin{figure}[t]
    \centering
    \includegraphics[width=\columnwidth]{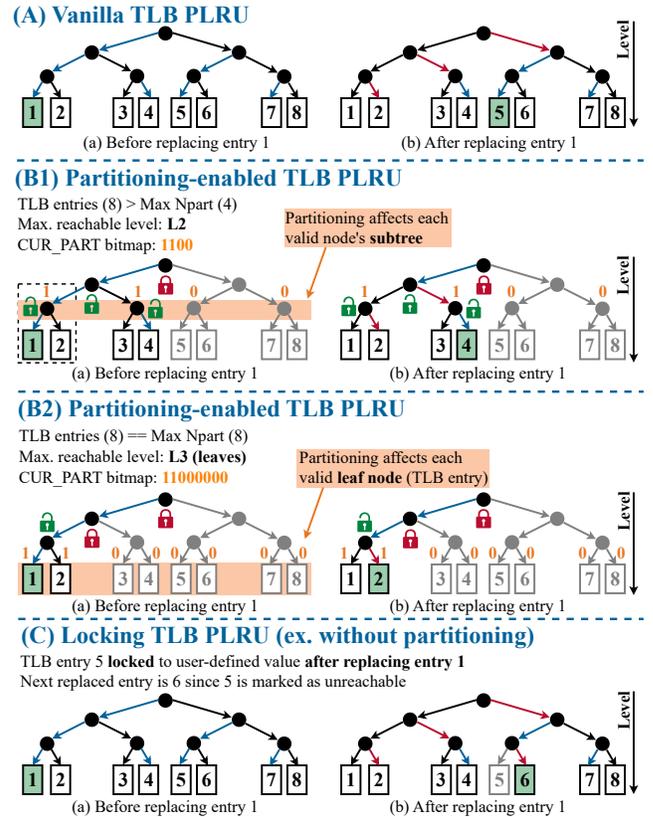}
    \caption{%
        PLRU behavior example for an eight-entry TLB.
    }
    \label{fig:archi:tlbpart}
\end{figure}

\subsection{Exemplary PLRU behavior}

To illustrate the partition-enabled \gls{plru} policy, figure~\ref{fig:archi:tlbpart} provides an example of an eight-entry \gls{tlb} with a visual representation of the corresponding \gls{plru} tree that would be used in CVA6. In all cases, we show the state before (a) and after (b) replacing the entry at index 1. The blue edges show the path from the tree root to the next entry to be evicted. The red edges illustrate how \gls{plru} updates the tree to select the next \gls{tlb} entry for replacement by flipping the edges along the chosen path.

In figure~\ref{fig:archi:tlbpart} A, the default case is shown, which results in entry number 5 being marked as the next victim by the tree.
In B1 and B2, we show our partition-enabled \gls{plru} tree with control over the tree's branching possibilities.
For B1, we show the case where the number of \gls{tlb} entries is larger than the number of partitions, meaning a single bit in \texttt{CUR\_PART} controls access to an entire sub-tree. In contrast, in B2 we have a one-to-one mapping of partitions to \gls{tlb} entries, as the number of entries and partitions match. In both cases, we observe that \gls{tlb} entries associated with cleared bits in the \texttt{CUR\_PART} \gls{csr} are protected from replacement by being unreachable from the tree root.
Case C highlights how the behavior of the non-partitioned \gls{plru} tree changes if we lock \gls{tlb} entry number 5. After replacing entry number 1, the tree would generally point to entry number 5. However, once a locking becomes valid, the associated leaf node is transparently marked as unreachable, independently of the state of the \texttt{CUR\_PART} \gls{csr}. Therefore, the entry that will be replaced next in case C is entry number 6. As shown in figure~\ref{fig:eval:cheshire}, \gls{tlb} locking and partitioning can cooperate to improve virtual-to-physical memory address translations of concurrent mixed-critical tasks.

\subsection{SPM extended cache controllers}
\label{sec:spm}
To further extend our system's timing predictability, we enhance the L1 data and instruction caches with a hybrid cache/\gls{spm} mode. This extension trades a part of the cache memory for software-controlled \gls{spm}, providing core-local memory with constant access time that is not influenced by hardware.

We achieve this by selectively removing cache ways from the associativity available to the cache replacement logic and reusing these memory macros for the \gls{spm} (figure~\ref{fig:archi:cva6vmrt}). Beyond modifying the way selection logic, we ensure that the associated tags and valid bits are cleared so that accidental cache hits on \gls{spm} data are impossible.
To distinguish between \gls{spm} and other memory accesses, we extend the cache controllers with address decoding logic that selects the correct destination, given the physical address of the memory access. This way, the \gls{spm} regions can be mapped into the standard address space seen by the processor core.

The cache ways are mapped contiguously in the \gls{spm} region, simplifying the mapping from the physical address to the corresponding cache way in hardware.
Before accessing the memory, the hardware ensures the target cache way is configured as \gls{spm}.
If the way is not configured correctly, write requests are silently dropped, and read requests or instruction fetches respond with dummy data to avoid stalling the processor.

By utilizing this hybrid cache/\gls{spm} mode, the \gls{os} can ensure that essential data and instructions, for example, critical interrupt handlers and their associated interrupt stacks, are always available with minimal latency, minimizing the overall interrupt handler latency.

\section{Evaluation}
\label{sec:eval}


\begin{figure}[t]
    \centering
    \includegraphics[width=\columnwidth]{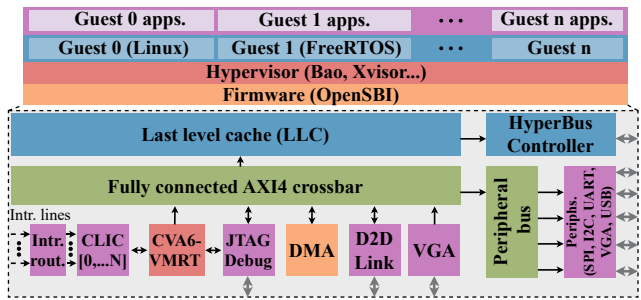}
    \caption{%
        Cheshire microarchitecture and software stack. 
    }
    \label{fig:eval:cheshire}
\end{figure}

\begin{figure*}[t]
    \centering
    \subfloat[No \gls{spm} used.]{
        \includegraphics[width=.97\columnwidth]{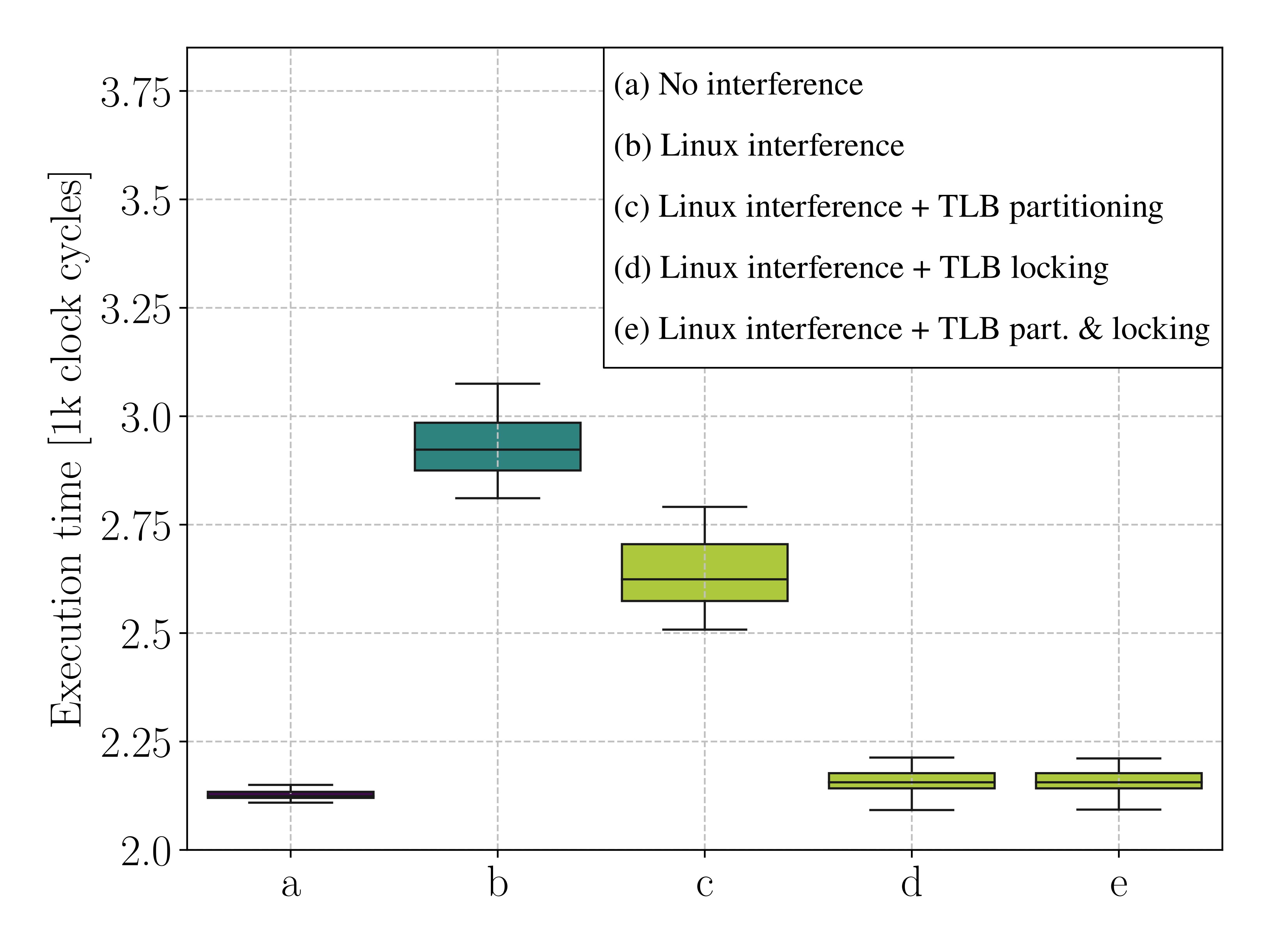}
        \label{fig:eval:rtos-nospm}
    }
    \hfill
    \subfloat[Using SPM.]{
        \includegraphics[width=.97\columnwidth]{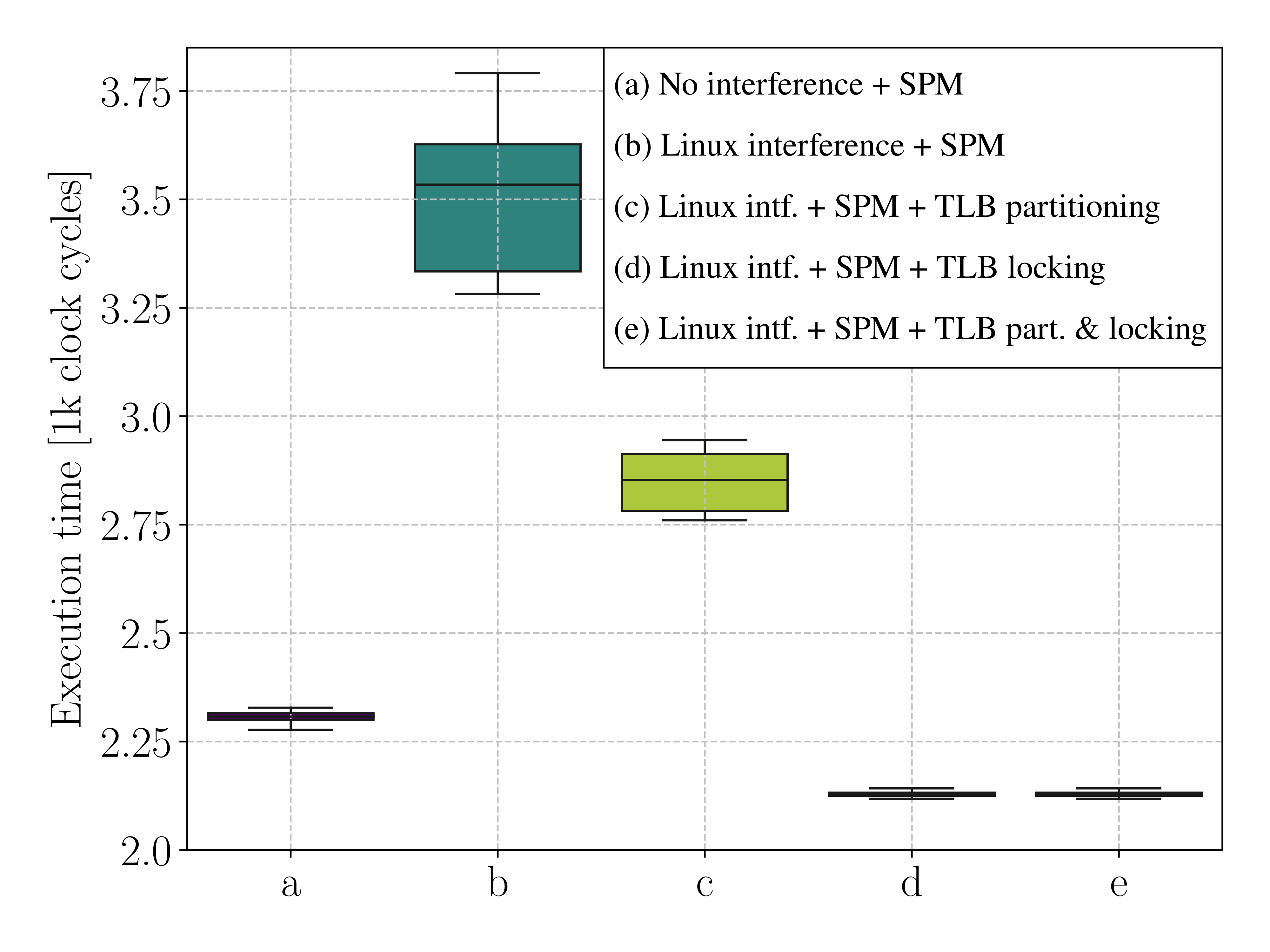}
        \label{fig:eval:rtos-spm}
    }
    \vspace{2mm}
    \caption{%
        Box plots for the synthetic benchmark using different interference mitigation configurations. Without our \gls{spm} extensions, \gls{tlb} locking achieves a standard deviation reduction of \SI{60}{\percent} compared to the unmitigated case. Using \gls{spm} additionally to locking further decreases the standard deviation, bringing the total reduction to \SI{91}{\percent}.
    }
    \label{fig:eval:cva6vmrt-synthetic}
\end{figure*}

\begin{figure}[t]
    \centering
    \includegraphics[width=\columnwidth]{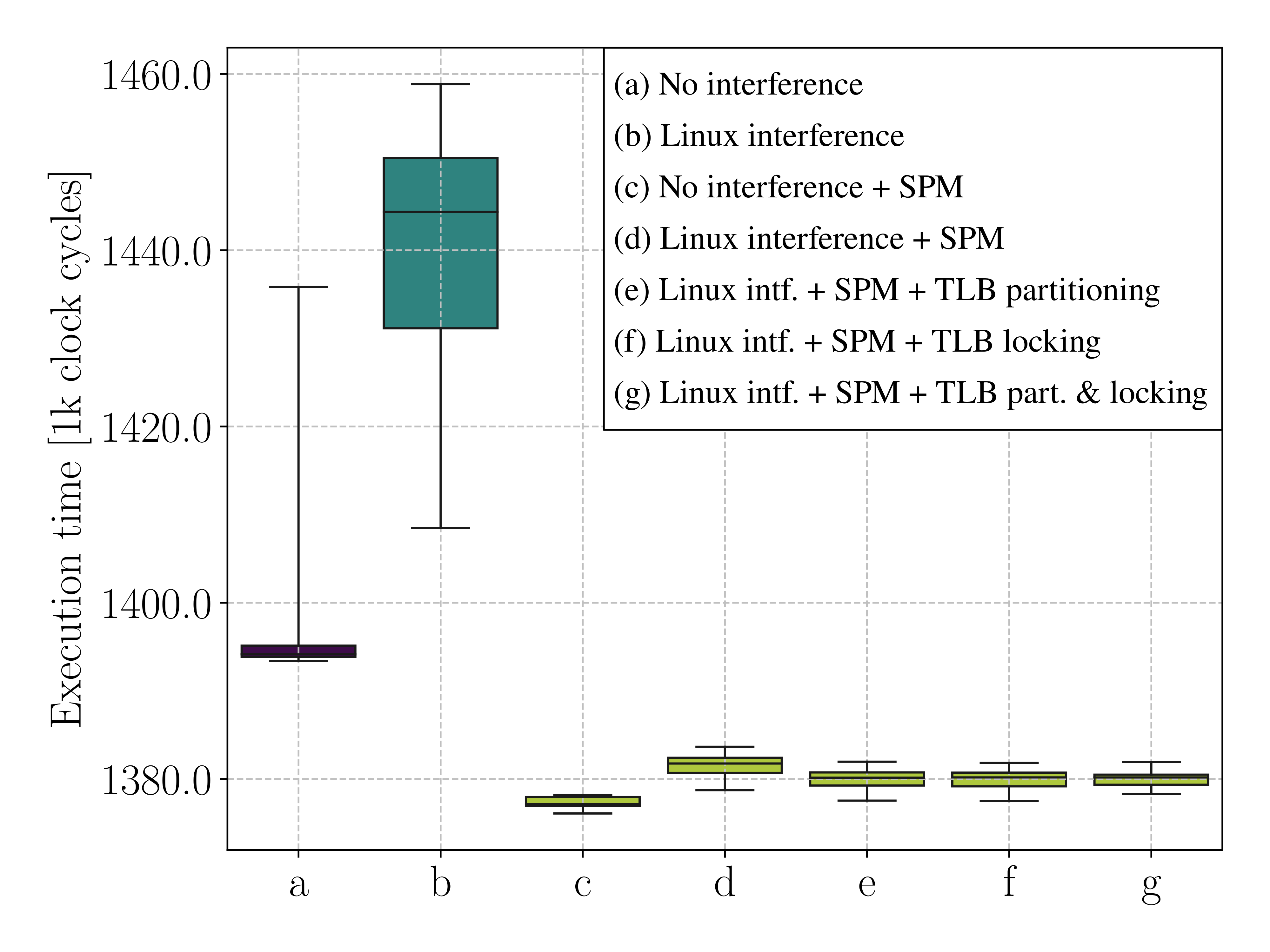}
    \caption{%
        Results of the \emph{powerwindow} benchmark for varying levels of enabled interference mitigations. Combining \gls{tlb} partitioning and locking with the \gls{spm} functionality \emph{CVA6-VMRT} achieves a \SI{94}{\percent} reduction in execution time standard deviation over the unmitigated case.
    }
    \label{fig:eval:cva6vmrt-app}
\end{figure}

\begin{figure}[t]
    \centering
    \includegraphics[width=\columnwidth]{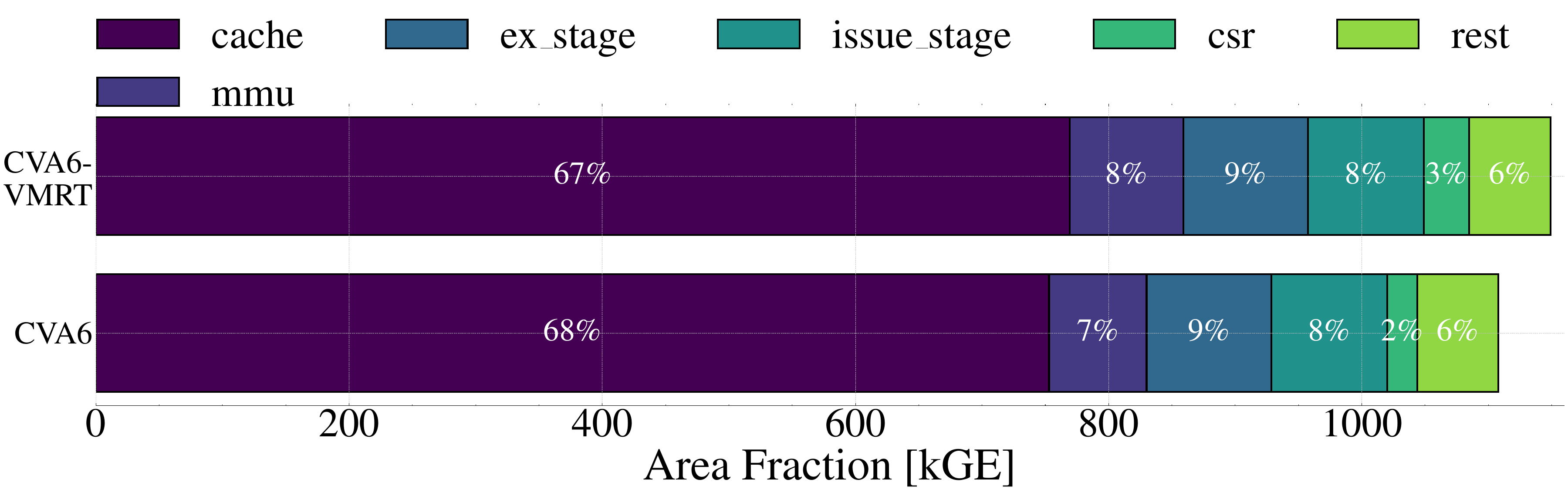}
    \caption{%
        CVA6-VMRT area breakdown. The overall area increase compared to vanilla CVA6 is about \SI{3.7}{\percent}.
    }
    \label{fig:eval:cva6vmrt-area}
\end{figure}

\subsection{Evaluation Framework}

We use the minimal Linux-capable host platform Cheshire~\cite{10163410} in a virtualized environment using Xvisor~\cite{xvisor} as \gls*{vmm} to evaluate our hardware modifications. Figure~\ref{fig:eval:cheshire} shows Cheshire's block diagram and software stack with virtualization support.
To collect a statistically significant set of measurements, we use the Digilent Genesys II \gls{fpga} implementation of Cheshire, extended with \emph{CVA6-VMRT}.

\paragraph{Interference Setup}
The scenario we consider is inspired by the trend of merging multiple different \glspl{ecu} into a single zonal controller using virtualization in modern automotive systems. 
It consists of two virtual machines running atop Xvisor on a single \emph{CVA6-VMRT} core. One of these \glspl{vm} executes a timing-critical task. At the same time, the other virtual machine hosts a Busybox +  Linux v6.1.64 environment running a synthetic memory-intensive process. This process uses as many \gls{tlb} entries as possible by continuously accessing different virtual memory pages. We use the Linux \gls{vm} as \emph{noise source} of our system to evaluate the isolation capabilities of our hardware extensions.

\paragraph{Time-critical Benchmarks}
We consider two types of time-critical tasks: representative benchmarks from \gls{sota} suites and synthetic benchmarks. The latter are crucial for assessing the extended hardware features in a controlled, streamlined environment.
As synthetic benchmark, we use a minimal bare-metal application designed to access the same number of virtual memory pages as there are \gls{tlb} entries.
As a representative benchmark, we select \textit{powerwindow} from the TACLeBench\cite{TACLeBench} suite, which models the control system of powered windows in modern vehicles. This benchmark represents control-intensive real-time tasks commonly found in automotive platforms, such as \glspl{adas} or engine control systems.

\paragraph{Xvisor Extensions}
We extend the Xvisor virtual machine configuration to enable support for the \gls{tlb} partitioning and locking mechanisms. 
This allows to specify a set of \gls{tlb} partitions on a per \gls{vm} basis as well as marking individual memory regions of each \gls{vm} as \gls{tlb} locked.
\Gls{tlb} locking is only supported when the targeted memory region is naturally aligned in both the physical and virtual address spaces and when their size is a multiple of the base page size or one of the SV39 super page sizes. 
A new locking entry is used for each required page table entry. In hardware, it is possible to add locking support to all \gls{tlb} entries; however, we chose to configure Xvisor to support at most eight locked entries as that is sufficient for our evaluation cases and minimizes the code repetition, which is necessary to access the different \glspl{csr} for each entry.
The creation of locked memory regions is handled during the initial \gls{vm} creation and uses a first-come, first-served basis, so it is up to the user to allocate locked entries to \glspl{vm}.

To handle \gls{tlb} partitioning, Xvisor requires minimal changes to its trap handler to save and restore the active partitions. With the first instruction in the interrupt handler, we overwrite the \texttt{CUR\_PART} \gls{csr} with a compile-time constant partition reserved for the hypervisor to minimize \gls{tlb} disturbance. The last instruction of the handler uses the \texttt{RESTORE\_LAST\_PART} register to atomically restore the value from \texttt{LAST\_PART} into \texttt{CUR\_PART}. If the trap did not cause a rescheduling of \glspl{vm}, this restores the partition register to the state before the trap was taken. However, we also modify the \gls{vm} switching code path to set the \texttt{LAST\_PART} \gls{csr} to the \gls{tlb} partition bitmap configured for the \gls{vm} that is scheduled next. This way, the last instruction of the trap handler switches to the new set of allowed \gls{tlb} partitions when the currently active \gls{vm} is changed.  

\paragraph{Evaluated Configurations}
In the test scenarios that utilize \gls{tlb} partitioning, we reserve one-half of the \gls{plru} tree for the critical \gls{vm}, a single entry of the other half for Xvisor itself, and the remainder for the Linux \gls{vm}. If \gls{tlb} locking is used, we use \gls{tlb} entries from the Linux \gls{vm} to store the locked entries as it is the less critical \gls{vm}.

For the tests using the hybrid \gls{spm}, we statically partition the available cache memory into $50\%$ cache and $50\%$ \gls{spm} during boot, exposing the \gls{spm} regions as regular memory to Xvisor, staying entirely transparent to the \glspl{vm}. We choose this ratio to balance the predictability improvement for the \gls{vm} using \gls{spm} with the impact of the reduced cache size on the non-\gls{spm} \gls{vm}. This trade-off should be carefully considered for each real-life application.
We allocate as much data and code as possible to the available \gls{spm} sections. In the case of the \textit{powerwindow} benchmark, we cannot fit the entire data set into the \gls{dspm} without changing the 50\% split ratio, so we keep the ratio unchanged and prioritize large continuous data structures for allocation into the \gls{dspm}, filling the remaining space with smaller variables.

\paragraph{Performance Metrics}

The primary performance metrics are the mean and standard deviation of the clock cycles required to complete a time-critical task under three scenarios: isolation, unregulated \gls{tlb} interference, and regulated \gls{tlb} interference with the proposed hardware extensions. The smaller the standard deviation, the higher the execution time predictability of the critical task.
To evaluate inter-VM interference, the synthetic benchmark primes the \glspl{tlb} without timing, then measures execution after de- and re-scheduling by Xvisor, accessing previously primed virtual memory pages in reverse order to avoid self-eviction of \gls{tlb} entries.

Experiments are conducted over \si{10000} iterations for all scenarios, analyzing the impact on the mean and standard deviation of execution time (section~\ref{subsec:eval:functional}). Hardware overhead is assessed by synthesizing Cheshire in a FinFET Intel \SI{16}{\nano\meter} technology node (section~\ref{subsec:eval:area}).

\subsection{Functional results}\label{subsec:eval:functional}

In figure~\ref{fig:eval:rtos-nospm}, we show the results for the synthetic benchmark without utilizing our hybrid cache/\gls{spm} functionality.
The X-axis shows the \gls{tlb} interference mitigations in use and whether the interference \gls{vm} is running. 
The Y-axis reports the distribution of the benchmark's execution time as a multiple of \SI{1000}{} clock cycles.

Scenario (a) marks the best-case scenario, with no other \gls{vm} active in the system. In figure~\ref{fig:eval:rtos-nospm}-(b), the Linux interference \gls{vm} is enabled, causing a \SI{38}{\percent} increase in average execution time and a standard deviation increase of \SI{652}{\percent}.
In figure~\ref{fig:eval:rtos-nospm}-(c), we partition the data and instruction \glspl{tlb} in the three non-overlapping sections mentioned above.
This partitioning reduces the impact of the interference on the mean execution time by \SI{36}{\percent}, but at the same time, further increases the standard deviation by \SI{10}{\percent}.\\
This issue is addressed by scenarios (d) and (e), which add locked \gls{tlb} entries, covering the entire memory space of the benchmark \gls{vm}. This reduces the mean execution time overhead over the isolated case to \SI{1.4}{\percent}. The remaining difference is due to other unmitigated interference sources like the caches or the memory controller. By utilizing \gls{tlb} locking, we achieve a decrease in the standard deviation of around \SI{60}{\percent} compared to the unmitigated case.\\
We also plot the combined use of \gls{tlb} partitioning and locking for completeness. However, the result remains within the margin of error compared to the locking-only configuration, as the locking mechanism already handles all necessary memory translations. As this might not be possible in all cases, for example in scenarios where the hypervisor is dynamically creating and tearing down virtual machines, potentially using fragmented physical memory to back a continuous region of memory in a \gls{vm}, we still see the case for \gls{tlb} partitioning, as it protects against external interference without assumptions on the virtual-to-physical memory mapping.

The same sequence of mitigations is applied to figure~\ref{fig:eval:rtos-spm}, with the only difference being the use of the hybrid \gls{spm} for the main benchmark routine, where most execution time is concentrated.
The first three configurations show between \SI{8}{\percent} and \SI{19}{\percent} increased mean execution times compared to their counterparts that do not use the \gls{spm} in figure~\ref{fig:eval:rtos-nospm}.
This happens because the benchmark routine is placed in a separate code section, limiting applicable compiler optimizations and reducing code locality.\\
Additionally, the \gls{spm}-enabled benchmark requires an extra \gls{pte} for the \gls{spm} region.
However, once the virtual-to-physical translation is provided statically from a locked \gls{tlb} entry, the average execution time falls to the level of the isolated case without \gls{spm} use (figure~\ref{fig:eval:rtos-nospm}, scenario 1). When the complete set of \emph{CVA6-VMRT} extensions is utilized (i.e., \gls{tlb} locking in combination with \gls{spm} usage), we see a decrease in execution time standard deviation of around \SI{91}{\percent} compared to the unmitigated case that is not using \gls{spm}.

For the \emph{powerwindow} benchmark, we combine both the results without and using the hybrid \gls{spm} functionality in figure~\ref{fig:eval:cva6vmrt-app} to highlight the cumulative isolation effect of our extensions. The box plot (a) shows the isolated case in which only the benchmark \gls{vm} is present in the system without mitigations activated.\\
Configuration (b) introduces the Linux interference \gls{vm}, which causes a \SI{3}{\percent} increase in the mean and a \SI{23}{\percent} increase in the standard deviation of the execution time. In scenario (c), we add our hybrid L1 cache/\gls{spm} functionality to scenario (a) and move the application code and most of its data into the private \gls{spm} partitions.\\
This marks the best-case scenario for the benchmark, as most data resides in non-evictable memory, and the remaining data can only be evicted by hypervisor interference.\\
In scenario (d), we re-introduce the interference from the Linux \gls{vm}, which now causes a mean execution time increase of \SI{0.3}{\percent} and a standard deviation increase of \SI{89}{\percent}. Compared to the isolated case without \gls{spm} (a), the use of \gls{spm} reduced the standard deviation by \SI{92}{\percent}.
Nonetheless, the following three scenarios show that interference through the \glspl{tlb} still harms the standard deviation of the execution time, as our \gls{tlb} partitioning (e), \gls{tlb} locking (f), and the combination of both (g) manage to reduce it further.\\
In combination, our extensions achieve a reduction of execution time standard deviation of around \SI{94}{\percent} for the \emph{powerwindow} benchmark.


Our findings show that \emph{CVA6-VMRT} can significantly reduce the standard deviation of the execution time experienced by virtualized critical tasks in the presence of interference by other \glspl{vm}. This increases the timing predictability of the protected critical tasks. In our benchmarks, we have the case that the \gls{spm}-backed code and data are always part of the working set.\\
Therefore, the average execution time is reduced, as this statically allocates a part of the working set to fast memory, which cannot be evicted. If this is not the case, the mean execution time may increase due to the reduced cache resources available.\\
This trade-off should be carefully considered when deciding what data and instructions should be placed in \gls{spm} and how much cache space can be sacrificed.

\subsection{Physical implementation}\label{subsec:eval:area}
To evaluate the area overhead of the changes introduced by \emph{CVA6-VMRT}, we synthesize identically configured CVA6s, both with and without our extensions, using an Intel 16 flow at \SI{600}{\mega\hertz}. We use the RCSS (slow) technology corner at \SI{0.72}{\volt} and \SI{125}{\celsius}. In both cases, the instruction cache is configured to be \SI{16}{\kibi\byte} large and the data cache \SI{32}{\kibi\byte}. The instruction and data \glspl{tlb} are configured to hold 16 entries each.

Figure~\ref{fig:eval:cva6vmrt-area} provides an area breakdown in kilo gate equivalents (kGE) of the two configurations.
The two main hardware units that grew in size are the \gls{mmu}, which contains both \glspl{tlb}, and the caches. In both cases, the area increase is only caused by the control logic and \glspl{csr} we added, as we reuse all memory macros. The \gls{mmu} incurs an overhead of \SI{16.4}{\percent}, and the caches of \SI{2.2}{\percent}. Our extensions did not noticeably affect the timing.
Compared to the whole CVA6 processor core, all changes require a mere \SI{3.7}{\percent} of additional area, which we consider negligible compared to the gained feature set.

\section{Related Work}
\label{sec:relwork}



We focus on techniques or existing processors that try to achieve deterministic virtual-to-physical address translation and deterministic instruction and data execution for critical tasks.
For this, previous works explored multiple approaches to increase \gls{tlb} and cache content predictability.
Table~\ref{tab:related_work} summarizes other works and compares them on their control over \gls{tlb} resources and use of \gls{spm}.

Panchamukhi \emph{et al.} extend the coloring technique from caches to the \glspl{tlb} in \cite{7108391}. By controlling the virtual addresses returned from \emph{malloc}, it is possible to ensure that the corresponding \gls{pte} entries do not collide in the set associative L2 \gls{tlb}, enabling \gls{tlb} partitioning without hardware overhead. This, however, relies on \gls{os} support and does not map well to our virtual machine host case.
With \emph{DTLB}~\cite{7884780} Varma \emph{et al.} propose a modified \gls{tlb} architecture that allows to create and restore backups of the \gls{tlb} state. To increase the memory translation time predictability of tasks preempted by an \gls{os}, the \gls{tlb} entries are saved as part of the task's state and restored when the task is rescheduled. This partitions the \glspl{tlb} in time instead of space, retaining the number of available \gls{tlb} entries while still providing isolation across processes. However, without hardware support for \gls{tlb} entry locking, processes can still suffer from the timing variability caused by intra-process \gls{tlb} interference.

A different approach to timing predictable virtual memory translation is taken by the Infineon TriCore~\cite{tricore_mmu}, which does not support virtual memory using page tables. Instead, it exposes software-managed memory map segments, which the \gls{os} configures with the currently active memory address translations. This means address translations have a constant latency or cause a page fault exception. While this achieves the same as our \gls{tlb} locking, it severely limits the system's usability, as the \gls{os} has to manage the range of currently accessible virtual memory on a very fine-grained basis.
Additionally, the TriCore features separate \glspl{spm} that coexist with the caches to provide predictable and low latency instruction and data access.

To increase the timing predictability of critical tasks, software-managed memories have been employed in ARM's Cortex-R82AE~\cite{cortex_r82ae_ref_manual}, which adds separate memory macros for \gls{spm}. These \glspl{spm} are meant for critical code sections and data, so fast and predictable access latencies can be guaranteed. However, they can only be accessed from a \gls{pmsa} context and are not usable when ARM's \gls{vmsa} is active, forbidding memory address translation when accessing \glspl{spm}. The intention is to provide a high-performance, 64-bit real-time processor that seamlessly integrates into a heterogeneous \gls{soc} while supporting rich \glspl{os}. With multiple R82AE cores in a system, the trade-off between rich \gls{os} and real-time compute power can then be managed dynamically by re-assigning cores to different \glspl{os}. The limitation is the physical-only \gls{spm} addressing, making it harder to use in virtualized contexts. Our implementation allows for full virtual or physical addressing without incurring the considerable area overhead of completely separate memory macros.

Similarly, to reduce the memory access time variability during instruction fetches, Cilku \emph{et al.} rely on single-path code, prefetching, and cache locking in their work on time-predictable instruction-cache architecture~\cite{7160126}. By exploiting the regular structure and simpler control flow in single-path code, the proposed prefetcher is guided on what instructions should be fetched sequentially and where loop prefetching is necessary. This way, given the absence of external interrupts, it is possible to provide deterministic instruction access times. Unfortunately, modern applications and execution environments do not fulfill these requirements, and fully software-managed memories like our hybrid L1 cache/\gls{spm} are more manageable.

The Gaisler LEON 5~\cite{gaisler_leon5_ref_manual} core supports tightly coupled memories, which can be added additionally to the default cache memories. Regardless of the page table contents, they are mapped to a specific virtual address. To provide isolation, the accessibility can be restricted to a single \gls{mmu} context, or access can be granted to all contexts simultaneously.  While this allows the use of virtual memory on \gls{spm}, it is a limitation compared to the arbitrary mappings supported by our implementation and still incurs the area overhead of separate memory macros. Additionally, the core provides a mechanism to freeze cache contents upon reception of an asynchronous interrupt. In this mode, the cache contents are kept coherent with the main memory, but no cache lines are evicted or replaced, preserving the set of cache lines present in the cache before the interrupt was taken. This is an effective way to retain predictability for a single protected thread but is not beneficial in a multi-process environment with multiple critical processes, as only one thread can be protected from interference by other threads. Additionally, the granularity of protection is the entire cache. In contrast, our \gls{spm} can be made available to arbitrary numbers of threads at a granularity of a single cache way.

The heterogeneous PolarFire \gls{soc} from Microchip~\cite{microchip_polarfire_soc_ref_manual} consists of one physical addressing only monitor core and four application class cores that support the RISC-V SV39 virtual memory specification. The monitor core features a software-managed \gls{dspm} instead of a data cache and supports reconfiguration of up to 50\% of its \SI{16}{\kibi\byte} instruction cache into \gls{ispm}. The application class cores do not include the \gls{dspm} and instead rely on a standard L1 data cache. Their instruction cache retains the hybrid cache/\gls{ispm} functionality and allows for up to \SI{28}{\kibi\byte} of \gls{ispm}. This memory can hold arbitrary data and instructions, but data load and store operations are less efficient than the monitor core's dedicated \gls{dspm}. Compared to our unified approach, one limitation of this approach is that the predictability and feature set are not uniform across all available cores. The monitor core can provide the highest predictability but no virtual memory. In contrast, the application cores can support a virtualized environment but suffer from less efficient data accesses to the \gls{ispm} and worse overall memory access time predictability due to the unmitigated memory address translation. This reduces the system's flexibility and increases the complexity of the programming model.

The Arm Cortex-A8~\cite{cortex_a8_ref_manual} core offers both \gls{tlb} lockdown and L2 cache lockdown. Using \gls{tlb} lockdown, it is possible to retain the contents of certain \gls{tlb} entries to ensure constant address translation latency. In secure privileged modes, it is furthermore possible to manually specify the contents of arbitrary L1 \gls{tlb} entries. The processor's L2 cache lockdown mechanism provides the user control over cache replacements at the granularity of cache ways. With this approach, L2 cache ways can effectively be used as \gls{spm} by locking all other cache ways and pre-loading the desired data by accessing each corresponding cache line, finally inverting the locking to disable evictions of the pre-loaded data. This ensures critical data and instructions are kept in the L2 cache, achieving the same predictability as an L2 \gls{spm} without needing a separate memory macro.
While this feature set is comparable to ours, it differs in key points. The cache lockdown feature is only available in the second cache level, meaning the user has to compromise on performance or predictability by disabling or enabling the L1 caches, respectively. Another key difference is the usability aspect, as the locked cache ways are not directly writeable but are pre-loaded using cache misses. This also requires care to ensure no conflict misses in the critical data or code.
Our implementation avoids these complications by providing direct load and store access to the hybrid L1 \gls{spm}.

\begin{table}[t]
    \begin{center}
    \setlength{\tabcolsep}{1.5pt} 
    \renewcommand{\arraystretch}{1.3} 
    \resizebox{\linewidth}{!}{
    \begin{tabular}{lccccc}

    \toprule

    \multirow{2}{*}{\rotatebox{0}{\textbf{Reference}}} &
    \multirow{2}{*}{\rotatebox{0}{\textbf{\makecell[cc]{Virtual \\ memory support}}}} & 
    \multirow{2}{*}{\rotatebox{0}{\textbf{\makecell[cc]{Predictable \\ core-local memory}}}} &
    \multirow{2}{*}{\rotatebox{0}{\textbf{\makecell[cc]{\gls{tlb} \\ control}}}} & 
    \multirow{2}{*}{\rotatebox{0}{\textbf{\makecell[cc]{Control \\ approach}}}} & 
    \multirow{2}{*}{\rotatebox{0}{\textbf{\makecell[cc]{Memory \\ macro reuse}}}} \\
    \\
    \midrule

    \textbf{Arm Cortex-A8~\cite{cortex_a8_ref_manual}} & Armv7 VMSA & L2 cache locking \& preloading & \ding{51} & HW & \ding{51}\\
    \\

    \textbf{Arm Cortex-R82AE~\cite{cortex_r82ae_ref_manual}} & Armv8 VMSA & \multirow{2}{*}{\makecell[cc]{\glspl{spm} \\ (physical addressing only)}} & \ding{55} & - & \ding{55} \\
    \\

    \textbf{Gaisler LEON 5~\cite{gaisler_leon5_ref_manual}} & SPARC V8 & \multirow{2}{*}{\makecell[cc]{Cache freezing \& \\ optional \glspl{spm} (physical addressing only)}} & \ding{55} & - & \ding{51} \ding{55}\\
    \\

    \textbf{Infineon TriCore~\cite{tricore_mmu}} & \multirow{2}{*}{\makecell[cc]{Custom \\ (\gls{tlb} only)}} & \gls{spm} & \ding{51} & HW & \ding{55} \\
    \\
    \textbf{Microchip PolarFire SoC~\cite{microchip_polarfire_soc_ref_manual}} & RISC-V SV39 & Partial hybrid cache/\gls{spm} & \ding{55} & - & \ding{51} \\
    \\
    \midrule
    \textbf{Panchamukhi \emph{et al.}~\cite{7108391}} & Required & \textit{n/a} & \ding{51} & SW & - \\

    \textbf{Varma \emph{et al.}~\cite{7884780}} & Required & \textit{n/a} & \ding{51} & HW & - \\

    \textbf{Cilku \emph{et al.}~\cite{7160126}} & \textit{n/a} & \multirow{2}{*}{\makecell[cc]{SW-manged I\$ w/ \\ prefetching \& locking}} & - & - & \ding{51} \\
    \\
    \midrule
    \textbf{This work} & \textbf{RISC-V SV39x4} & \textbf{Hybrid L1 cache/\gls{spm}} & \ding{51} & \textbf{HW} & \ding{51}\\

    \bottomrule
    \end{tabular}

    }
    \vspace{2mm}
    \caption{Comparison of approaches to increased system predictability focusing on the virtual memory system and core-local memories.}
    \label{tab:related_work}
    \end{center}
    \vspace{-7mm}
\end{table}

To the best of our knowledge, no state-of-the-art work combines a predictable virtual memory system with runtime configurable hybrid L1 cache/\gls{spm} partitions to reduce variability in critical virtual memory accesses on RISC-V.
\section{Conclusion}
\label{sec:concl}
This paper presents \emph{CVA6-VMRT}, an enhanced version of the open-source RISC-V processor CVA6, which facilitates dynamic, software-controlled partitioning of \gls{tlb} resources and enables the runtime configurable trade-off of cache space for deterministic \glspl{spm} to improve predictable execution of time-critical tasks in non-federated, virtualized \glspl*{mcs}. In experiments with two virtual machines competing for \gls{tlb} and cache resources, \emph{CVA6-VMRT} successfully reduced execution time variability of the protected virtual machine by \SI{91}{\percent} in synthetic benchmarks and \SI{94}{\percent} in application-class benchmarks. When implemented using a commercial \SI{16}{\nano\meter} technology node, \emph{CVA6-VMRT} introduced only a \SI{3.7}{\percent} area overhead compared to the baseline CVA6 processor, making it a cost-effective extension for \glspl{mcs} based on CVA6.

\begin{acks}
\label{sec:ack}
    %
This work has received funding from the Swiss State
Secretariat for Education, Research, and Innovation (SERI) under the
SwissChips initiative %
and
was partly supported through the ISOLDE project that has received funding from Chips Joint Undertaking (CHIPS-JU) under grant agreement nr. 101112274. CHIPS JU receives support from the European Union’s Horizon Europe’s research and innovation programme and Austria, Czechia, France, Germany, Italy, Romania, Spain, Sweden, Switzerland.

\end{acks}
\bibliographystyle{ACM-Reference-Format}
\bibliography{new_main}


\end{document}